\RequirePackage{fixltx2e}
\documentclass[a4paper]{isp}
\usepackage[american]{babel}
\usepackage{graphicx}
\usepackage{amsmath,amsthm,mathtools}
\usepackage{url}
\usepackage[capitalise,nameinlink]{cleveref}
\usepackage{xcolor}
\usepackage[utf8]{inputenc}
\usepackage{enumitem}

\def\apj{ApJ\,  }

\def\mnras{MNRAS\,  }

\def\sun{\hbox{$\odot$}}

\begin{document}
\pdfgentounicode=1
\title
{
New probability distributions in astrophysics:
V. The truncated Weibull distribution
}
\author{Lorenzo  Zaninetti}
\institute{
Physics Department,
 via P.Giuria 1, I-10125 Turin,Italy \\
 \email{zaninetti@ph.unito.it}
}

\maketitle

\begin {abstract}
We demonstrate that certain astrophysical distributions 
can be modelled with
the truncated Weibull distribution, which can lead to some insights:
in particular, we
report   
the average value,
the $r$th moment,
the variance,
the median,
the mode,
the  generation of random numbers, 
and
the evaluation of the two parameters with
maximum likelihood
estimators.
The first application  of the Weibull distribution
is to the initial  mass function for stars.
The magnitude version of the Weibull  distribution  is
applied  to the luminosity function for the Sloan Digital Sky Survey (SDSS)
 galaxies
and to the photometric maximum of the 2MASS Redshift Survey 
(2MRS) galaxies.
The truncated Weibull   luminosity  function  allows us
to model the average value of the absolute magnitude  
as a function of the redshift 
for the  2MRS galaxies.
\end{abstract}

\section{Introduction}
The Weibull distribution
was  originally introduced  
to model the 
fracture strength of brittle and quasi-brittle materials,
see   \cite{Weibull1939,Weibull1951}.
The Weibull distribution was successively applied 
to analyse 
the voltage breakdown of electric circuits
\cite{Cacciari1996,Hirose1996},
the  life table data of plants such as the 
Kolla paulula \cite{Tuan2017},
the  distribution of tree diameters \cite{Diamantopoulou2015}, 
plant vegetative tissue \cite{Pitt1982}
and  the 
fatigue failure studies of human extensor digitorum longus  
\cite{Gallagher2012}.
The analysis of the truncated Weibull distribution has  
been explored in many papers, we list some in what follows.
The upper truncated Weibull distribution 
has been analysed  
by  \cite{Zhang2011}
and applied to modeling component or system failure,
and 
by \cite{Kantar2015} 
to modeling wind speed data and estimating wind power density.
The lower truncated Weibull distribution 
has been analysed  
by  \cite{Wingo1989}.
The lower and upper truncated  Weibull distribution and the evaluation 
of its  moments  has been analysed in \cite{Mcewen1991,Khan2007,Crenin2015}.
A careful analysis of the above approaches allows concluding 
that the truncated  Weibull distribution 
has not yet  been applied to astrophysics.
This  paper  
reviews the existing Weibull 
distribution 
in Section \ref{section_review},
introduces the truncated  Weibull distribution
in Section \ref{section_truncated},
derives  the Weibull luminosity function  (LF)
for galaxies and the connected modification 
due to the truncation in Section \ref{section_luminosity},
and  
discusses,  in  Section \ref{section_astro}, 
the application of the Weibull LF 
to the SDSS galaxies, 
to QSOs,
to the photometric maximum visible in the 2MRS catalog of galaxies,
and to the  cosmological evolution
of the average absolute  magnitude with the redshift.

\section{The Weibull  distribution}
\label{section_review}

\label{weibullsec}
Let $X$ be a random variable
defined in
$[0, \infty]$;
the two/parameter {\em Weibull}
distribution function (DF), $F(x)$,
 is
\begin{equation}
F (x;b,c) = 
1-{{\rm e}^{- \left( {\frac {x}{b}} \right) ^{c}}}
\quad,
\end{equation}
where $b$ and $c$, both positive, are the scale and the 
shape parameters, see \cite{evans2011}.
The probability density function (PDF), $f(x)$,
is
\begin{equation}
f (x;b,c) =
\frac
{
c{x}^{c-1}{{\rm e}^{- \left( {\frac {x}{b}} \right) ^{c}}}
}
{
{b}^{c}
}
\quad .
\label{weibullpdfscaling}
\end{equation}
We now introduce the  function
\begin{equation}
\Gamma_i = \Gamma (1 +i/c)
\quad ,
\end{equation}
the average  value or mean, $\mu$, is
\begin{equation}
\mu (b,c)=
b\Gamma_{{1}}
\quad  ,
\end{equation}
the variance, $\sigma^2$, is
\begin{equation}
\sigma^2(b,c)=
{b}^{2} \left( -{\Gamma_{{1}}}^{2}+\Gamma_{{2}} \right) 
\quad ,
\end{equation}
the skewness is
\begin{equation}
skewness(b,c)=
{\frac {2\,{\Gamma_{{1}}}^{3}-3\,\Gamma_{{2}}\Gamma_{{1}}+\Gamma_{{3}}
}{ \left( {\Gamma_{{1}}}^{2}-\Gamma_{{2}} \right) ^{2}}\sqrt {-{\Gamma
_{{1}}}^{2}+\Gamma_{{2}}}}
\quad ,
\end{equation}
and the kurtosis
\begin{equation}
kurtosis(b,c)=
-{\frac {3\,{\Gamma_{{1}}}^{4}-6\,\Gamma_{{2}}{\Gamma_{{1}}}^{2}+4\,
\Gamma_{{1}}\Gamma_{{3}}-\Gamma_{{4}}}{ \left( -{\Gamma_{{1}}}^{2}+
\Gamma_{{2}} \right) ^{2}}}
\quad  .
\end{equation}

The 
$r$th moment about the origin for the Weibull distribution,
$\mu^{\prime}_r$, 
is
\begin{equation}
\mu^{\prime}_r(b,c) =
{b}^{r}\Gamma \left( {\frac {c+r}{c}} \right)
\quad ,
\end{equation}
where $r$ is an integer 
and  
\begin{equation}
\mathop{\Gamma\/}\nolimits\!\left(z\right)
=\int_{0}^{\infty}e^{{-t}}t^{{z-1}}dt
\quad ,
\end{equation}
is the gamma function, see \cite{NIST2010}.
The median    is at
\begin{equation}
{{\rm e}^{{\frac {\ln  \left( \ln  \left( 2 \right)  \right) +c\ln 
 \left( b \right) }{c}}}}
\quad ,
\end{equation}
and the mode is at
\begin{equation}
\sqrt [c]{{\frac {c-1}{c}}}b
\quad .
\end{equation}
Random generation of the Weibull  variate $X$ 
is given by
\begin{equation}
X:b,c \approx 
\sqrt [c]{-\ln  \left( 1-R \right) }b
\end{equation}
where 
$R$ is the unit rectangular variate.
The two parameters $b$ and $c$ can be derived
by the numerical solution  of the two following  equations
which arise from the maximum likelihood estimator (MLE)
\begin{subequations}
\begin{align}
{\frac {c}{b} \left( \sum _{i=1}^{n} \left( {\frac {x_{{i}}}{b}}
 \right) ^{c}-n \right) }=0 \quad , \\
-n\ln  \left( b \right) +{\frac {n}{c}}+\sum _{i=1}^{n}- \left( {
\frac {x_{{i}}}{b}} \right) ^{c}\ln  \left( {\frac {x_{{i}}}{b}}
 \right) +\ln  \left( x_{{i}} \right) =0
\quad  ,
\end{align}
\end{subequations}
where $x_i$ are the elements  of the experimental sample 
with  $i$  varying between  1 and  $n$.

\section{The truncated Weibull  distribution}

\label{section_truncated}
Let $X$ be a random variable
defined in
$[x_l,x_u]$;
the truncated two-parameter {\em Weibull}
DF, $F_T(x)$,
 is
\begin{equation}
F_T (x;b,c,x_l,x_u) = 
\frac
{
-{{\rm e}^{- \left( {\frac {x}{b}} \right) ^{c}}}+{{\rm e}^{- \left( {
\frac {x_{{l}}}{b}} \right) ^{c}}}
}
{
-{{\rm e}^{- \left( {\frac {x_{{u}}}{b}} \right) ^{c}}}+{{\rm e}^{-
 \left( {\frac {x_{{l}}}{b}} \right) ^{c}}}
}
\quad ,
\end{equation}
and 
the PDF, $f_T(x)$,
is
\begin{equation}
f_T (x;b,c,xl,xu) =
\frac
{
- \left( {\frac {x}{b}} \right) ^{c}c{{\rm e}^{- \left( {\frac {x}{b}}
 \right) ^{c}}}
}
{
x \left( {{\rm e}^{- \left( {\frac {x_{{u}}}{b}} \right) ^{c}}}-{
{\rm e}^{- \left( {\frac {x_{{l}}}{b}} \right) ^{c}}} \right) 
}
\quad ,
\label{pdfweibulltruncated}
\end{equation}
see  Section 2.1 in \cite{Crenin2015}.

The inequality which fixes the range of existence
is $\infty>x_u>x>x_l>0$.
We report the indefinite integral which 
characterizes
the average  value or mean, $\mu_T$,
\begin{equation}
I(b,c,x_l,x_u,x) = \int x\,f_T (x;b,c)\, dx
\quad ,
\end{equation}
which is
\begin{equation}
I(b,c,x_l,x_u,x)
=
\frac{IN}{ID}
\quad ,
\end{equation}
where
\begin{eqnarray}
IN=
-2\,c\sqrt {x}{{\rm e}^{-\frac{1}{2}\,{b}^{-c}{x}^{c}+c \left( -\ln  \left( x
 \right) +\ln  \left( b \right)  \right) }}{b}^{\frac{1}{2}-c} 
\nonumber
\\
\left( 2\,
 \left( \frac{1}{2}+c \right) ^{2}{b}^{c}{{\sl M}_{\frac{1}{2}\,{\frac {2\,c+1}{c}},\,
\frac{1}{2}\,{\frac {3\,c+1}{c}}}\left({b}^{-c}{x}^{c}\right)}+c \left( 
 \left( \frac{1}{2}+c \right) {b}^{c}+\frac{1}{2}\,c{x}^{c} \right) {{\sl M}_{\frac{1}{2}\,{c}
^{-1},\,\frac{1}{2}\,{\frac {3\,c+1}{c}}}\left({b}^{-c}{x}^{c}\right)}
 \right) 
\quad ,
\end{eqnarray}
and
\begin{equation}
ID =
\left( 1+c \right)  \left( 2\,c+1 \right)  \left( 3\,c+1 \right) 
 \left( {{\rm e}^{-{x_{{u}}}^{c}{b}^{-c}}}-{{\rm e}^{-{x_{{l}}}^{c}{b}
^{-c}}} \right) 
\end{equation}
where ${{\sl M}_{\mu,\,\nu}\left(z\right)}$ is 
the Whittaker $M$ function,
see  \cite{Abramowitz1965,NIST2010}.
The average value is therefore
\begin{equation}
\mu(b,c,x_l,x_u)=
I(b,c,x_l,x_u,x=x_u)
-
I(b,c,x_l,x_u,x=x_l)
\quad ,
\end{equation}
for a comparison, see equation~(5) in 
\cite{Crenin2015}.
The
indefinite integral 
which characterizes
the   
$r$th moment about the origin for the truncated Weibull distribution,
$\mu^{\prime}_{r,t}$, 
is
\begin{equation}
M(b,c,x_l,x_u,x) = \int x^r\,f_T (x;b,c)\, dx
\quad ,
\end{equation}
which is
\begin{equation}
M(b,c,x_l,x_u,x) =
\frac
{
MN
}
{
\left( -{{\rm e}^{-{x_{{u}}}^{c}{b}^{-c}}}+{{\rm e}^{-{x_{{l}}}^{c}{b
}^{-c}}} \right)  \left( c+r \right)  \left( 2\,c+r \right)  \left( 3
\,c+r \right)
}
\quad ,
\end{equation}
where
\begin{eqnarray}
MN=
2\,c{{\rm e}^{-1/2\,{b}^{-c}{x}^{c}+c   ( -\ln    ( x   ) +
\ln    ( b   )    ) }} 
\Bigg  ( 2\,{x}^{r/2}{b}^{r/2}   ( 
c+r/2   ) ^{2}{{\sl M}_{1+1/2\,{\frac {r}{c}},\,3/2+1/2\,{\frac {r
}{c}}}  ({b}^{-c}{x}^{c}  )}
\nonumber \\
+{{\sl M}_{1/2\,{\frac {r}{c}},\,3/
2+1/2\,{\frac {r}{c}}}  ({b}^{-c}{x}^{c}  )}c  \Big ( 1/2\,c{x}^
{c+r/2}{b}^{r/2-c}+{b}^{r/2}{x}^{r/2}   ( c+r/2   )    
\Big    ) 
\Bigg   ) 
\quad .
\end{eqnarray}
The   
$r$th moment about the origin for the truncated Weibull distribution
is therefore 
\begin{equation}
\mu^{\prime}_{r,t}=
M(b,c,x_l,x_u,x=x_u) -
M(b,c,x_l,x_u,x=x_l)
\quad .
\end{equation} 
The  variance, $\sigma^2_T(b,c,x_l,x_u)$,
of the truncated Weibull distribution
is given by
\begin{equation}
\sigma^2_T(b,c,x_l,x_u) =
 \mu^{\prime}_{2,t} -(\mu^{\prime}_{1,t})^2
\quad .
\end{equation}
The $median_T$    
in the case 
$x_u> median_T >x_l$
is at
\begin{equation}
median_T= 
\sqrt [c]{ \left( {x_{{u}}}^{c}+{x_{{l}}}^{c} \right) {b}^{-c}-\ln 
 \left( {\frac {{{\rm e}^{{x_{{u}}}^{c}{b}^{-c}}}}{2}}+{\frac {{
{\rm e}^{{x_{{l}}}^{c}{b}^{-c}}}}{2}} \right) }b
\quad ,
\end{equation}
and the $mode_T$ 
in the case 
$x_u> mode_T >x_l$
is at
\begin{equation}
mode_T=\sqrt [c]{{\frac {c-1}{c}}}b
\quad ,
\end{equation}
which is the same value as that for the Weibull pdf.
Random generation of the truncated Weibull  variate $X$ 
is given by
\begin{equation}
X:b,c,x_l,x_u \approx 
\sqrt [c]{ \left( {x_{{u}}}^{c}+{x_{{l}}}^{c} \right) {b}^{-c}-\ln 
 \left( -R{{\rm e}^{{x_{{u}}}^{c}{b}^{-c}}}+R{{\rm e}^{{x_{{l}}}^{c}{b
}^{-c}}}+{{\rm e}^{{x_{{u}}}^{c}{b}^{-c}}} \right) }b
\quad ,
\end{equation}
where 
$R$ is the unit rectangular variate.
The four parameters $x_l$, $x_u$, $b$ and $c$ 
can be obtained in the following way. 
Consider a  sample  ${\mathcal X}=x_1, x_2, \dots , x_n$ and let
$x_{(1)} \geq x_{(2)} \geq \dots \geq x_{(n)}$ denote
their order statistics, so that
$x_{(1)}=\max(x_1, x_2, \dots, x_n)$, $x_{(n)}=\min(x_1, x_2, \dots, x_n)$.
The first two parameters $x_l$ and $x_u$
are
\begin{equation}
{x_l}=x_{(n)}, \qquad { x_u}=x_{(1)}
\quad  .
\label{eq:firstpar}
\end{equation}
The MLE is obtained by maximizing
\begin{equation}
\Lambda = \sum_i^n \ln(f_T (x;b,c,xl,xu)).
\end{equation}
The two derivatives $\frac{\partial \Lambda}{\partial b} =0$ and
$\frac{\partial \Lambda}{\partial c}=0 $  generate two
non-linear equations in
 $b$ and $c $ which 
are 
\begin{subequations}
\begin{align}
\frac
{
N1
}
{
 \left( -{{\rm e}^{- \left( {\frac {x_{{u}}}{b}} \right) ^{c}}}+{
{\rm e}^{- \left( {\frac {x_{{l}}}{b}} \right) ^{c}}} \right) b
}
=0 \quad , \\
\frac
{
N2
}
{
\left( -{{\rm e}^{- \left( {\frac {x_{{u}}}{b}} \right) ^{c}}}+{
{\rm e}^{- \left( {\frac {x_{{l}}}{b}} \right) ^{c}}} \right) c
}
 =0
\quad  ,
\end{align}
\end{subequations}
where 
\begin{eqnarray}
N1=
 \Bigg (  \left( -{{\rm e}^{- \left( {\frac {x_{{u}}}{b}} \right) ^{c}}
}+{{\rm e}^{- \left( {\frac {x_{{l}}}{b}} \right) ^{c}}} \right) \sum 
_{i=1}^{n} \left( {\frac {x_{{i}}}{b}} \right) ^{c}
\nonumber \\
+ \left(  \left( -
 \left( {\frac {x_{{l}}}{b}} \right) ^{c}-1 \right) {{\rm e}^{-
 \left( {\frac {x_{{l}}}{b}} \right) ^{c}}}+{{\rm e}^{- \left( {\frac 
{x_{{u}}}{b}} \right) ^{c}}} \left(  \left( {\frac {x_{{u}}}{b}}
 \right) ^{c}+1 \right)  \right) n \Bigg) c
\quad  ,
\end{eqnarray}
and
\begin{eqnarray}
N2=
\left( -{{\rm e}^{- \left( {\frac {x_{{u}}}{b}} \right) ^{c}}}+{
{\rm e}^{- \left( {\frac {x_{{l}}}{b}} \right) ^{c}}} \right) c\sum _{
i=1}^{n}- \left( {\frac {x_{{i}}}{b}} \right) ^{c}\ln  \left( {\frac {
x_{{i}}}{b}} \right) +\ln  \left( x_{{i}} \right) +
\nonumber  \\
n \left(  \left( 
 \left( {\frac {x_{{l}}}{b}} \right) ^{c}\ln  \left( {\frac {x_{{l}}}{
b}} \right) c-c\ln  \left( b \right) +1 \right) {{\rm e}^{- \left( {
\frac {x_{{l}}}{b}} \right) ^{c}}}-{{\rm e}^{- \left( {\frac {x_{{u}}
}{b}} \right) ^{c}}} \left(  \left( {\frac {x_{{u}}}{b}} \right) ^{c}
\ln  \left( {\frac {x_{{u}}}{b}} \right) c-c\ln  \left( b \right) +1
 \right)  \right) 
\quad  .
\end{eqnarray}

\section{The luminosity function}

\label{section_luminosity}

This  section 
reports  the luminosity functions (LFs)
for   the Weibull distribution and
      the truncated Weibull distribution.

\subsection{The Weibull LF}

The  Schechter function,
introduced by
\cite{schechter},
provides a useful reference   for the
LF  of galaxies
\begin{equation}
\Phi (L;\alpha,L^*,\Phi^*) dL  = (\frac {\Phi^*}{L^*}) (\frac {L}{L^*})^{\alpha}
\exp \bigl ( {-  \frac {L}{L^*}} \bigr ) dL \quad  ,
\label{equation_schechter}
\end {equation}
here $\alpha$ sets the slope for low values
of $L$, $L^*$ is the
characteristic luminosity and $\Phi^*$ is the normalization.
The equivalent distribution in absolute magnitude is
\begin{equation}
\Phi (M)dM=0.921 \Phi^* 10^{0.4(\alpha +1 ) (M^*-M)}
\exp \bigl ({- 10^{0.4(M^*-M)}} \bigr)  dM \, ,
\label{lfstandard}
\end {equation}
where $M^*$ is the characteristic magnitude as derived from the
data.
We now  introduce  the parameter $h$,
which is $H_0/100$, where $H_0$ is the Hubble constant.
The scaling with  $h$ is  $M^* - 5\log_{10}h$ and
$\Phi^* ~h^3~[Mpc^{-3}]$.
In order to derive the  Weibull LF
we start from the PDF as given by
equation (\ref{weibullpdfscaling}),
\begin{equation}
\Psi(L;c,L^*,\Psi^*) dL  ={\it \Psi^*}
\frac
{
\left( {\frac {L}{{\it L^*}}} \right) ^{c}c{{\rm e}^{- \left( {
\frac {L}{{\it L^*}}} \right) ^{c}}}
}
{
L
}\, dL   \quad ,
\label{weibull_lf}
\end{equation}
where
$L$  is the luminosity, $L^*$ is the
characteristic luminosity and $\Psi^*$ is the normalization
and  
the version in absolute magnitude is
\begin{equation}
\Psi(M;c,M^*,\Psi^*) dM  = 
0.4\,{\it \Psi^*}\,{10}^{ \left( - 0.4\,M+ 0.4\,{\it M^*} \right) c}c{{\rm e}^{-
{10}^{ \left( - 0.4\,M+ 0.4\,{\it M^*} \right) c}}}\,
\ln  \left( 10 \right) 
\, dM
\quad .
\label{lfweibull}
\end{equation}

\subsection{The truncated Weibull LF}

We start with the   truncated Weibull PDF with scaling as given by
equation (\ref{pdfweibulltruncated})
\begin{equation}
\Psi(L;c,L^*,\Psi^*,L_l,L_u) dL  =
\Psi^*
\frac
{
- \left( {\frac {L}{{\it L^*}}} \right) ^{c}c{{\rm e}^{- \left( {
\frac {L}{{\it L^*}}} \right) ^{c}}}
}
{
L \left( {{\rm e}^{- \left( {\frac {L_{{u}}}{{\it L^*}}} \right) ^{c
}}}-{{\rm e}^{- \left( {\frac {L_{{l}}}{{\it L^*}}} \right) ^{c}}}
 \right) 
}
\,dL
\quad  ,
\end{equation}
where
$L$  is the luminosity,  $L^*$ is the
characteristic luminosity,
$L_l$ is the lower boundary in luminosity,
$L_u$ is the upper boundary in luminosity,
and $\Psi^*$ is the normalization.
The magnitude version is
\begin{eqnarray}
\Psi(M;c,M^*,\Psi^*,M_l,M_u) dM  =
\nonumber \\
\Psi^*
\frac
{
- 0.4\, \left( {10}^{ 0.4\,{\it M^*}- 0.4\,M} \right) ^{c}c{{\rm e}
^{- \left( {10}^{ 0.4\,{\it M^*}- 0.4\,M} \right) ^{c}}}\, \left( \ln  \left( 2 \right) +\ln  \left( 5 \right) 
 \right) 
}
{
{{\rm e}^{- \left( {10}^{- 0.4\,M_{{l}}+ 0.4\,{\it M^*}} \right) ^{
c}}}-{{\rm e}^{- \left( {10}^{ 0.4\,{\it M^*}- 0.4\,M_{{u}}}
 \right) ^{c}}}
}
\,dM
\quad  ,
\end{eqnarray}
where $M$ is the absolute magnitude,
$M^*$   the characteristic magnitude,
$M_l$   the lower boundary in  magnitude,
$M_u$   the upper boundary in  magnitude and
 $\Psi^*$ is the normalization.
The mean theoretical absolute  magnitude, ${ \langle M \rangle }$,
can  be evaluated as
\begin{equation}
 { \langle M \rangle }
=
\frac
{
\int_{M_l}^{M_u} M \times \Psi(M;c,M^*,\Psi^*,M_l,M_u) dM
}
{
\int_{M_l}^{M_u}  \Psi(M;c,M^*,\Psi^*,M_l,M_u) dM
}
\quad .
\label{meanabsolutetruncated}
\end{equation}

\section{Astrophysical  applications}

\label{section_astro} 
This  section 
reviews the adopted statistics,
applies the truncated Weibull distribution to the initial mass 
function (IMF) for stars,  
models the LF for galaxies and QSOs,
explains the  photometric maximum
in the number of galaxies of the  2MRS,
and traces the cosmological evolution of the average absolute magnitude.

\subsection{Statistics}

The merit function $\chi^2$
is computed
according to the formula
\begin{equation}
\chi^2 = \sum_{i=1}^n \frac { (T_i - O_i)^2} {T_i},
\label{chisquare}
\end {equation}
where $n  $   is the number of bins,
      $T_i$   is the theoretical value,
and   $O_i$   is the experimental value represented
by the frequencies.
The theoretical  frequency distribution is given by
\begin{equation}
 T_i  = N {\Delta x_i } p(x) \quad,
\label{frequenciesteo}
\end{equation}
where $N$ is the number of elements of the sample,
      $\Delta x_i $ is the magnitude of the size interval,
and   $p(x)$ is the PDF  under examination.

A reduced  merit function $\chi_{red}^2$
is  given  by
\begin{equation}
\chi_{red}^2 = \chi^2/NF
\quad,
\label{chisquarereduced}
\end{equation}
where $NF=n-k$ is the number of degrees  of freedom,
$n$ is the number of bins,
and $k$ is the number of parameters.
The goodness  of the fit can be expressed by
the probability $Q$, see  equation 15.2.12  in \cite{press},
which involves the number of degrees of freedom
and $\chi^2$.
According to  \cite{press} p.~658, the
fit `may be acceptable' if  $Q>0.001$. 

The Akaike information criterion
(AIC), see \cite{Akaike1974},
is defined by
\begin{equation}
AIC  = 2k - 2  ln(L)
\quad,
\end {equation}
where $L$ is
the likelihood  function  and $k$  the number of  free parameters
in the model.
We assume  a Gaussian distribution for  the errors.
Then  the likelihood  function
can be derived  from the $\chi^2$ statistic
$L \propto \exp (- \frac{\chi^2}{2} ) $
where  $\chi^2$ has been computed by
eq.~(\ref{chisquare}),
see~\cite{Liddle2004}, \cite{Godlowski2005}.
Now the AIC becomes
\begin{equation}
AIC  = 2k + \chi^2
\quad.
\label{AIC}
\end {equation}

The Kolmogorov--Smirnov test (K--S),
see \cite{Kolmogoroff1941,Smirnov1948,Massey1951},
does not  require binning the data.
The K--S test,
as implemented by the FORTRAN subroutine KSONE in \cite{press},
finds
the maximum  distance, $D$, between the theoretical
and the astronomical  CDF
as well the  significance  level  $P_{KS}$,
see formulas  14.3.5 and 14.3.9  in \cite{press};
if  $ P_{KS} \geq 0.1 $,
the goodness of the fit is believable.

\subsection{The IMF for stars}

We tested the truncated Weibull distribution
on four samples of stars:
NGC 2362  (271 stars),
the young cluster NGC 6611 (207 stars),
the $\gamma$ Velorum  cluster (237 stars),
and the 
young cluster Berkeley 59 (420 stars),
for more details, see Section 5.2 of \cite{Zaninetti2020d}.
The results are presented in
Table \ref{chi2weibulltrunc} for  the 
truncated Weibull distribution with two parameters,
where the last column reports 
whether the results are better compared to the lognormal
distribution (Y) or worse (N).
Results  on the lognormal distribution
were reported in Table 1 in \cite{Zaninetti2020d}.

\begin{table}[ht!]
\caption
{
Numerical values of
$\chi_{red}^2$, AIC, 
probability $Q$,
$D$, 
the maximum distance between theoretical and observed DF,
and  $P_{KS}$, 
significance level,   in the K--S test of the 
truncated Weibull distribution with two parameters 
for  different  mass distributions.
The last column (LN) indicates 
an AIC lower (Y) or bigger (N) in respect
to the lognormal distribution.
The  number of  linear   bins, $n$, is 20.
}
\label{chi2weibulltrunc}
\begin{center}
\resizebox{16cm}{!}
{
\begin{tabular}{|c|c|c|c|c|c|c|c|}
\hline
Cluster     &parameters &  AIC    & $\chi_{red}^2$ & $Q$   
            &  D       &   $P_{KS}$  & LN    \\
\hline
NGC~2362    &b=0.726 ,c=2.2 , $x_l= 0.12$,$x_u=1.47$   &   
39.5  &  1.96        & 
 0.011  &   0.011    &  0.576 & N \\
NGC~6611    &b=0.483, c= 1.011 ,$x_l= 0.019$,$x_u=1.46$    &  
47.77  &  2.48          
&  $8.4\,10^{-4}$      & 0.059    & 0.45 & Y \\
$\gamma$~Velorum    &b=0.153 , c= 0.745  ,$x_l= 0.158$,$x_u=1.317$   &
31.24  &  1.45          &  0.107  &0.063    & 0.292 & Y \\
Berkeley~59 & b=0.347   , c=  1.143  ,$x_l= 0.16$,$x_u= 2.24$   &
83.71  &   4.73
& $9.74\,10^{-10}$  & 0.122     & $6.35\,10^{-6}$ &N \\
\hline
\end{tabular}
}
\end{center}
\end{table}

\begin{figure*}
\begin{center}
\includegraphics[width=6cm]{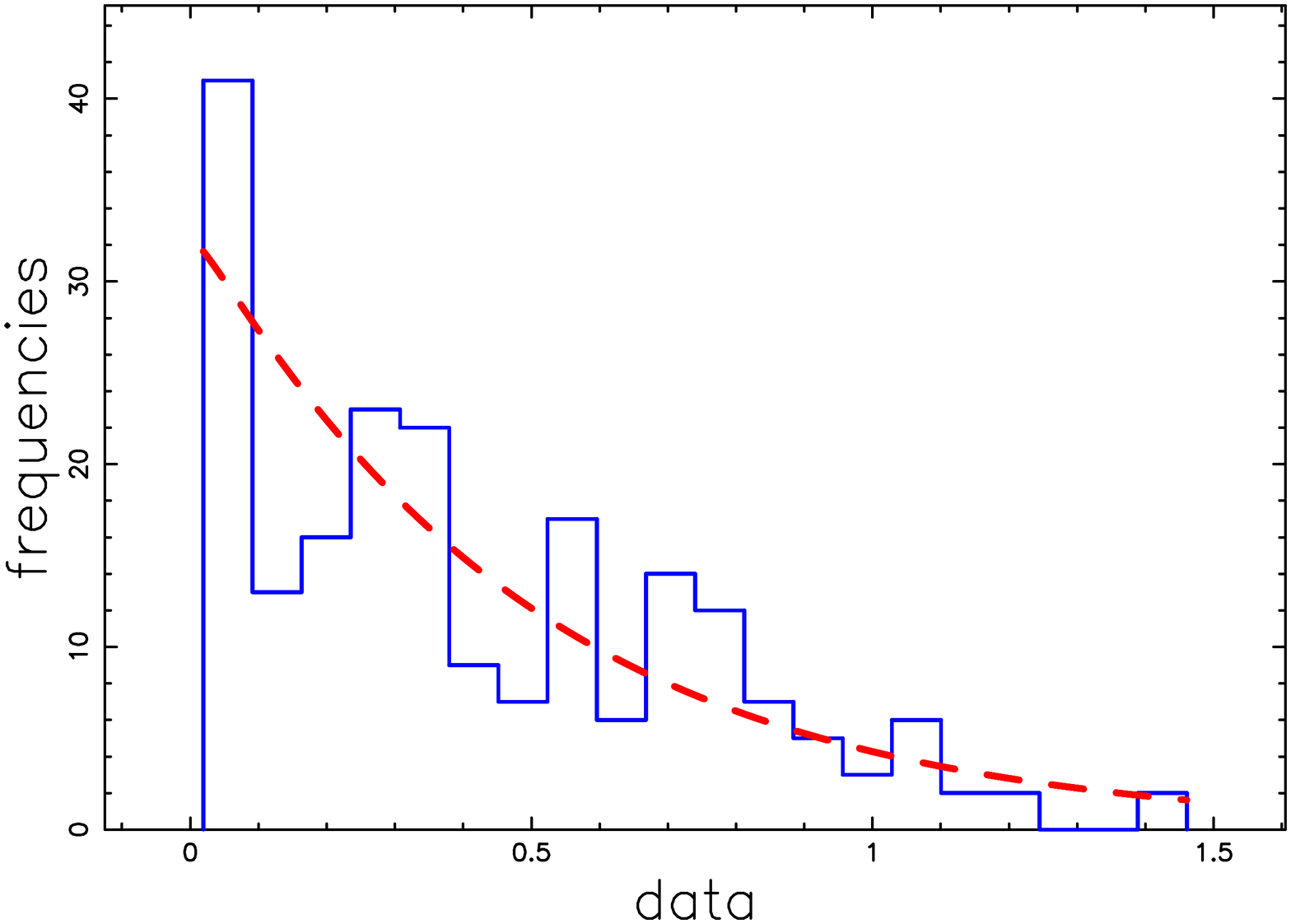}
\end{center}
\caption
{
Empirical  PDF   of  the mass distribution
for   NGC 6611 cluster data 
(blue histogram)
with a superposition of the truncated{} Weibull PDF 
(red  line).
Theoretical parameters as in Table \ref{chi2weibulltrunc}.
}
\label{weibull_ngc6611}
\end{figure*}

\begin{figure*}
\begin{center}
\includegraphics[width=6cm]{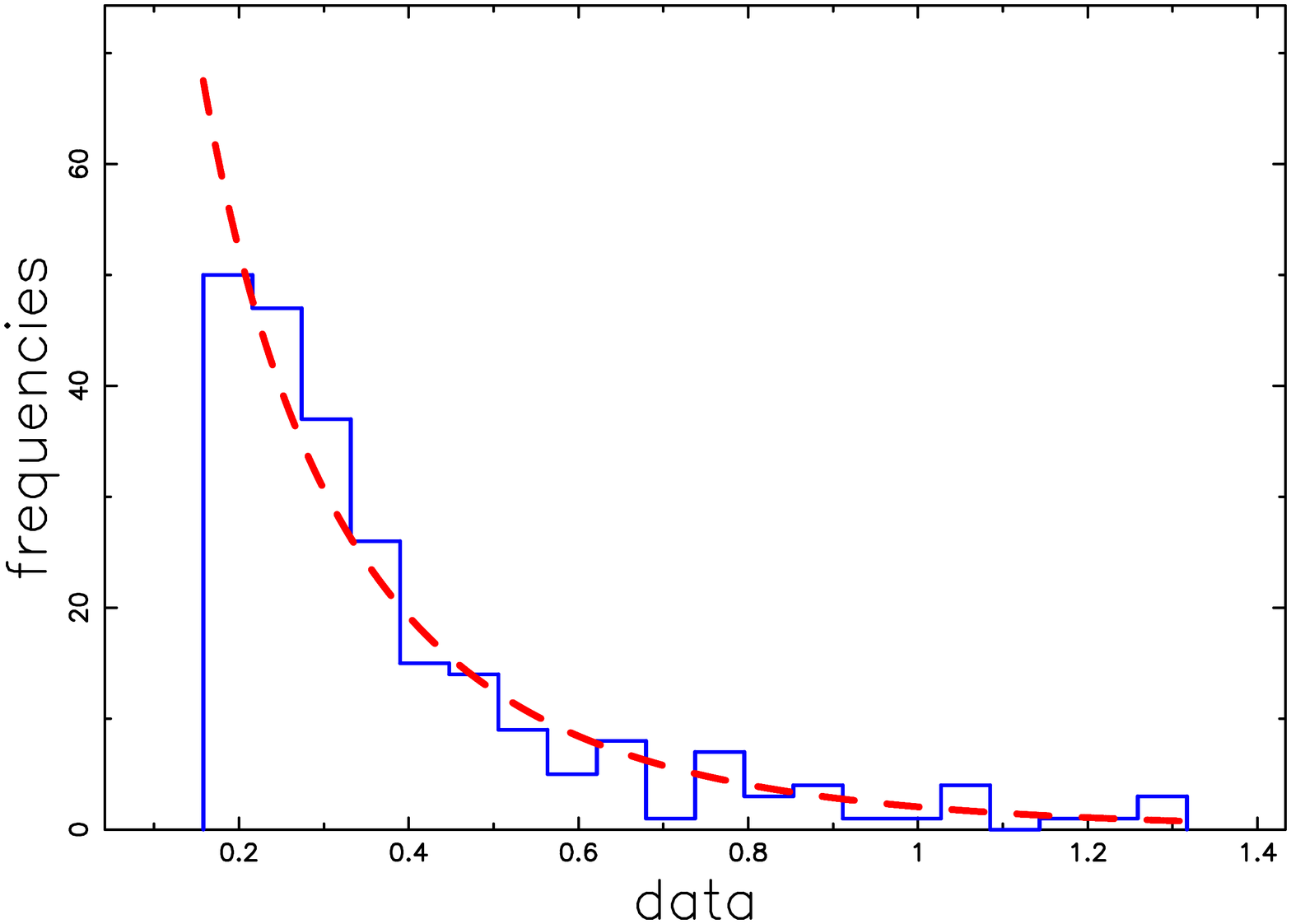}
\end{center}
\caption
{
Empirical  PDF   of  the mass distribution
for   $\gamma$~Velorum data 
(blue histogram)
with a superposition of the truncated{} Weibull PDF 
(red  line).
Theoretical parameters as in Table \ref{chi2weibulltrunc}.
}
\label{weibull_gamma_vel}
\end{figure*}

Graphical displays of the empirical PDF 
visualized through  histograms 
as well as the theoretical PDF 
for NGC 6611
are  reported in Figure \ref{weibull_ngc6611}
and 
those  for the $\gamma$~Velorum sample
are reported in Figure \ref{weibull_gamma_vel}.

\subsection{The LF for galaxies}

\begin{figure*}
\begin{center}
\includegraphics[width=10cm]{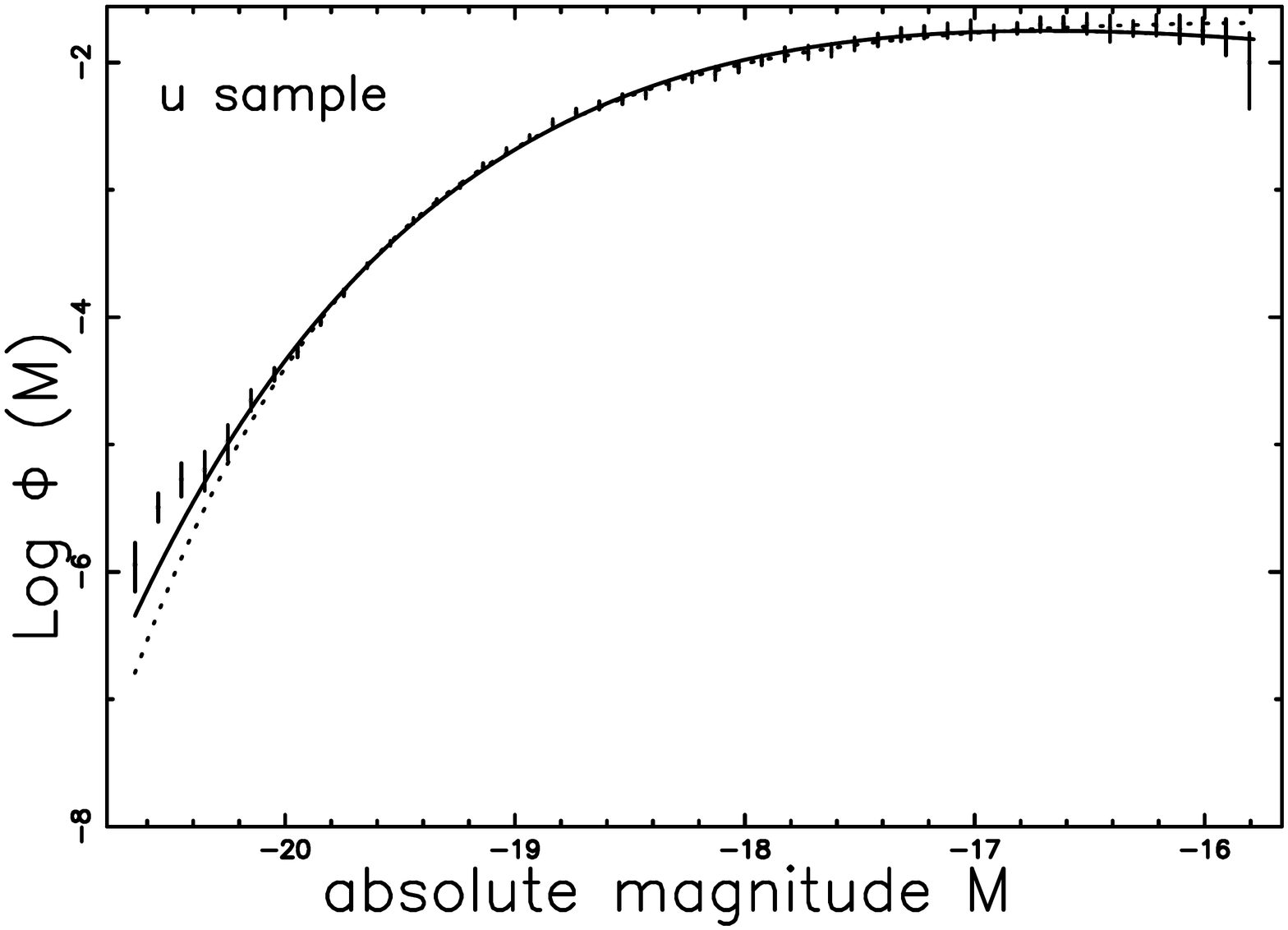}
\end{center}
\caption
{
The LF  data of
SDSS($u^*$) are represented with error bars.
The continuous line fit represents the Weibull LF
(\ref{lfweibull})
and the dotted
line represents the Schechter function.
}
\label{weibull_due_u}
\end{figure*}

A test has been  performed on the $u^*$
band  of SDSS as
in \cite{Blanton_2003} with data available
at \url{https://cosmo.nyu.edu/blanton/lf.html}.
The Schechter function, the new Weibull LF
represented by formula~(\ref{lfweibull})
and the data are
reported in
Figure~\ref{weibull_due_u}, parameters as
in Table \ref{chi2valuelf}.

\begin{table}[ht!]
\caption {
Numerical values and  $\chi^2_{red}$ of the LFs
applied to  SDSS Galaxies
in the $u^*$ band.
 }
\label{chi2valuelf}
\begin{center}
\begin{tabular}{|c|c|c|}
\hline
LF        &   parameters    & $\chi^2_{red}$ \\
\hline
Schechter &
$M^*$= -17.92 ,\, $\alpha$=-0.9,\, $\Phi^* = 0.03 /Mpc^3$
 & 0.689
  \\
Weibull &  $M^*$= -16.69 ,\, c=0.728 , \, $\Psi^* = 0.0718 /Mpc^3$
&   0.650   \\
\hline
\end{tabular}
\end{center}
\end{table}
A careful examination of Table \ref{chi2valuelf}
reveals that the Weibull LF has a lower $\chi_{red}^2$
compared to the Schechter LF.

\begin{figure}
\begin{center}
\includegraphics[width=7cm]{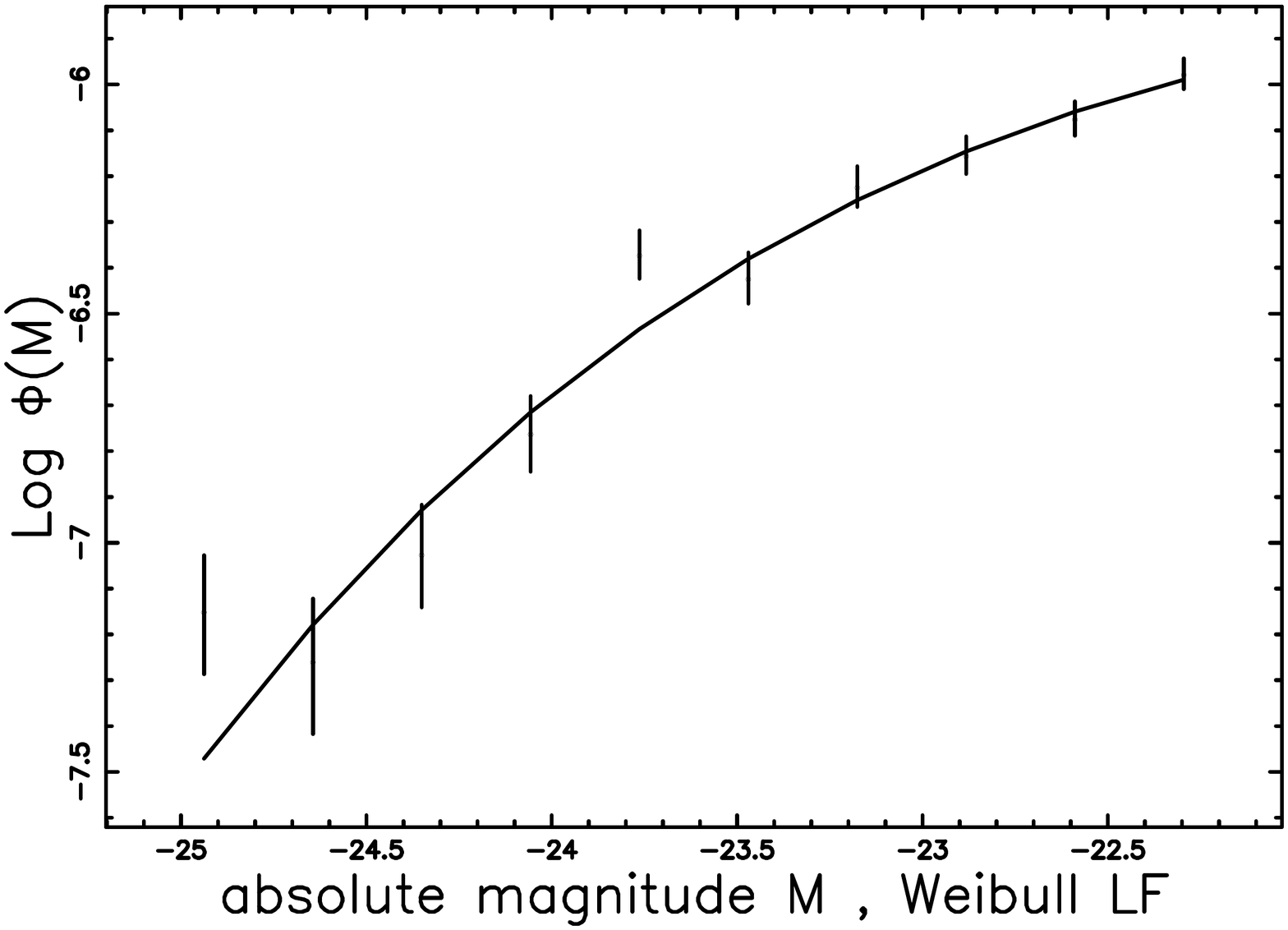}
\end{center}
\caption{
The  observed LF for QSOs, empty stars with error bar, 
and the fit  by  the Weibull  LF
for $z$ in $[0.3,0.5]$
and  $M$ in  $[-24.93,-22]$.
Parameters as in Table  \ref{qsoweibullfit}.
}
 \label{weibull_qso}%
\end{figure}

Another case is the LF for QSO in the case 
$ 0.3 < z< 0.5$, see  \cite{Zaninetti2017a}
for more details.
Figure \ref{weibull_qso} displays the 
observed LF for QSO as well the theoretical 
fit with the Weibull LF.
The parameters  and the statistical results
for the Schechter LF  are reported 
in Table \ref{schechterfit} 
and those 
for the Weibull LF  
in Table \ref{qsoweibullfit};
the Weibull LF  has smaller
$\chi_{red}^2$  compared to the 
Schechter LF.

\begin{table}
\caption
{
Parameters  of the  Schechter LF 
in the range of redshift $[0.3,0.5]$
when $k=3$ and $n=10$.
}
 \label{schechterfit}
 \[
 \begin{tabular}{ccccccc}
 \hline
 \hline
 \noalign{\smallskip}
$ M^*$
& $\Psi^*$
&  $\alpha$ 
&  $\chi^2$
&  $\chi_{red}^2$
& $Q$  
& AIC  
\\
 \noalign{\smallskip}
 \hline
 -23.75  & 8.85\,$10^{-7}$& -1.37 & 10.49 
& 1.49  & 0.162  & 16.49
\\
 \hline
 \hline
 \end{tabular}
 \]
\end{table}

\begin{table}
\caption
{
Parameters  of the  Weibull LF for QSOs  
in the range of redshift $[0.3,0.5]$
when $k=3$ and $n=10$.
}
 \label{qsoweibullfit}
 \[
 \begin{tabular}{ccccccc}
 \hline
 \hline
 \noalign{\smallskip}
 $M^*$
& $\Psi^*$
& c
&  $\chi^2$
& $\chi_{red}^2$
& $Q$  
& AIC  
\\
 \noalign{\smallskip}
 \hline
 -20.566  &  9.26\,$10^{-6}$  &  0.471 &  10.08
& 1.44  & 0.183  & 16.08
\\
 \hline
 \hline
 \end{tabular}
 \]
\end{table}

\subsection{The photometric maximum}

In the pseudo-Euclidean universe,
the  correlation
between the expansion velocity  and distance is
\begin {equation}
V= H_0 D  = c_l \, z
\quad  ,
\label {clz}
\end{equation}
where $H_0$ is the Hubble constant,
$H_0 = 100 h \mathrm{\ km\ s}^{-1}\mathrm{\ Mpc}^{-1}$, with $h=1$
when  $h$ is not specified,
$D$ is the distance in Mpc,
$c_l$ is  the  speed of light and
$z$
is the redshift.
In the
pseudo-Euclidean
universe,
the flux of radiation,
$f$,  expressed in units of $ \frac {L_{\sun}}{Mpc^2}$,
where $L_{\sun}$ represents the luminosity of the sun,  is
\begin{equation}
f  = \frac{L}{4 \pi D^2}
\quad ,
\end{equation}
where $D$   represents the distance of the galaxy
expressed in Mpc,
and
\begin{equation}
D=\frac{c_l z}{H_0}
\quad  .
\end{equation}
The joint distribution in {\it z}
and {\it f}  for  a generic  LF, $\Phi (\frac{z^2}{z_{crit}^2}$
is
\begin{equation}
\frac{dN}{d\Omega dz df} =
4 \pi  \bigl ( \frac {c_l}{H_0} \bigr )^5    z^4 \Phi (\frac{z^2}{z_{crit}^2})
\label{nfunctionzgeneric}
\quad ,
\end {equation}
where $d\Omega$, $dz$ and  $df$ represent
the differentials of
the solid angle,
the redshift, and the flux, respectively,
and 
\begin{equation}
 z_{crit}^2 = \frac {H_0^2  L^* } {4 \pi f c_l^2}
\quad 
\end{equation}
where  $L^*$ is the 
characteristic luminosity, 
for more details, see  \cite{Zaninetti2019a}.
The LF  is chosen to be the  
Schechter function, but different LFs can be tested,
for example, the Weibull LF.
The joint distribution in
$z$, $f$
and $\Omega$ 
for galaxies  for  the
Weibull LF, see equation (\ref{weibull_lf}),
is
\begin{equation}
\frac{dN(z;c,\Psi^*,z_{crit})}{d\Omega dz df} =
\frac
{
4\,{z}^{2}{c_{{l}}}^{5}{\it \Psi^*}\, \left( {\frac {{z}^{2}}{{z_{{{
\it crit}}}}^{2}}} \right) ^{c}c\pi\,{z_{{{\it crit}}}}^{2}{{\rm e}^{-
 \left( {\frac {{z}^{2}}{{z_{{{\it crit}}}}^{2}}} \right) ^{c}}}
}
{
{H_{{0}}}^{5}{\it L^*}
}
\quad .
\label{nfunctionzweibull}
\end{equation}

\begin{figure}
\begin{center}
\includegraphics[width=6cm]{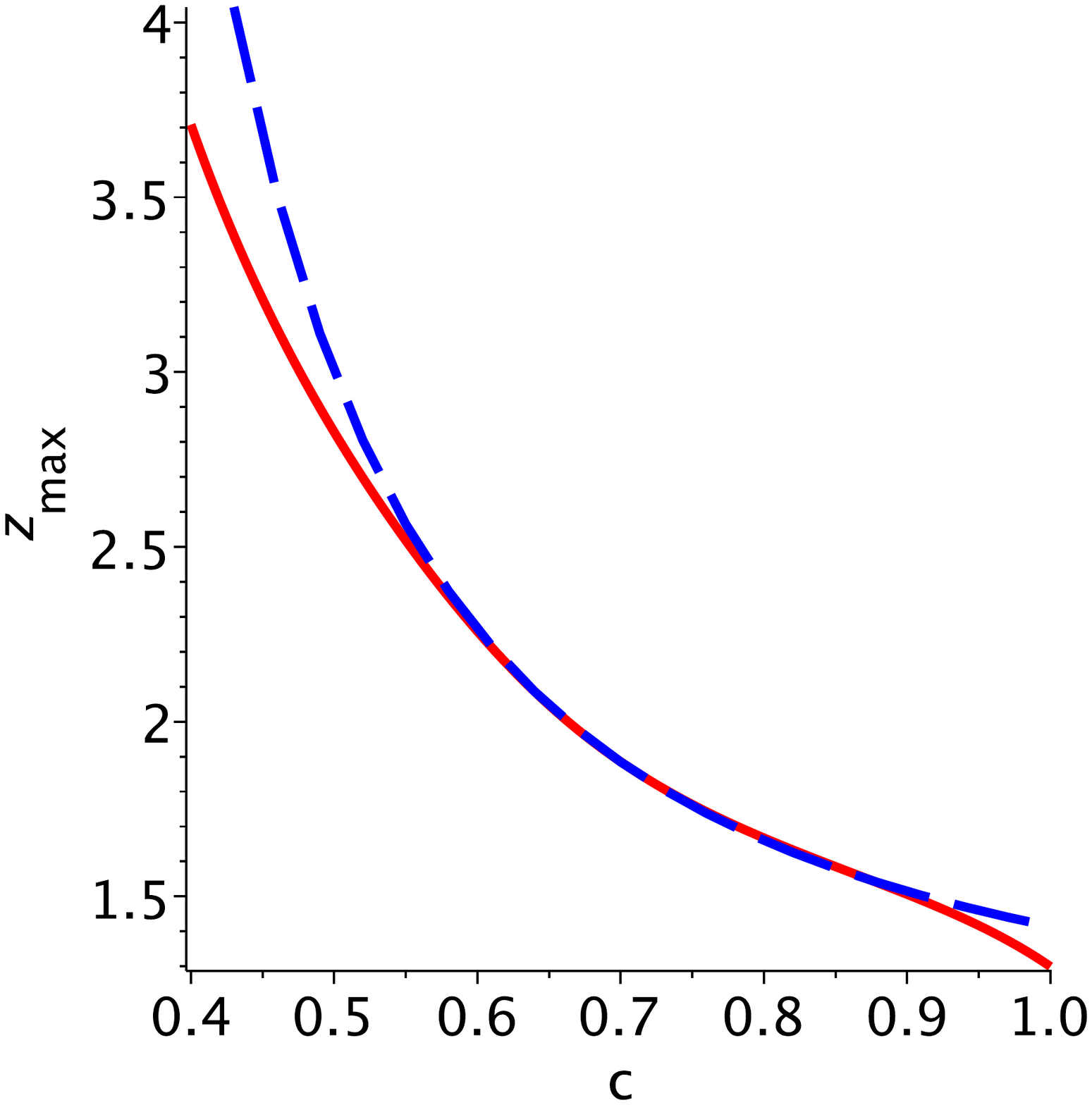}
\end {center}
\caption
{
Position in units of  $z_{crit}$ of the  photometric 
maximum as a function of the shape parameter $c$:
numerical solution (blue dashed line)
and
Taylor solution    (red full line).
}
          \label{zmaxtaylor}%
    \end{figure}

The above number of galaxies in $z$  and $f$ 
has a maximum  at  $z=z_{max}$
which is the solution  of the following non-linear
equation
\begin{equation}
-8\,z{c_{{l}}}^{5}{\it \Psi}\, \left( {\frac {{z}^{2}}{{z_{{{\it 
crit}}}}^{2}}} \right) ^{c}c\pi\,{z_{{{\it crit}}}}^{2}{{\rm e}^{-
 \left( {\frac {{z}^{2}}{{z_{{{\it crit}}}}^{2}}} \right) ^{c}}}
 \left(  \left( {\frac {{z}^{2}}{{z_{{{\it crit}}}}^{2}}} \right) ^{c}
c-c-1 \right) =0 
\quad .
\label{eqnozmax}
\end{equation}
A {\it first}  numerical evaluation
of the position in $z$ of the above equation  
is reported  in  units of $z_{crit}$,
see the blue dashed line
in Figure  \ref{zmaxtaylor}.
A {\it second} analytical result can be obtained 
inserting  for the number of galaxies a numerical value for 
$c$.
As an  example when $c=1/2$, the nonlinear equation 
for the photometric maximum is
\begin{equation}
-2\,z{c_{{l}}}^{5}{\it \Psi^*}\,\sqrt {{\frac {{z}^{2}}{{z_{{{\it 
crit}}}}^{2}}}}\pi\,{z_{{{\it crit}}}}^{2}{{\rm e}^{-\sqrt {{\frac {{z
}^{2}}{{z_{{{\it crit}}}}^{2}}}}}} \left( \sqrt {{\frac {{z}^{2}}{{z_{
{{\it crit}}}}^{2}}}}-3 \right) =0
\quad ,
\end{equation}
which has a physical solution
at
\begin{equation}
z_{max} = 3 z_{crit}
\quad  .
\end{equation}

\begin{figure}
\begin{center}
\includegraphics[width=6cm]{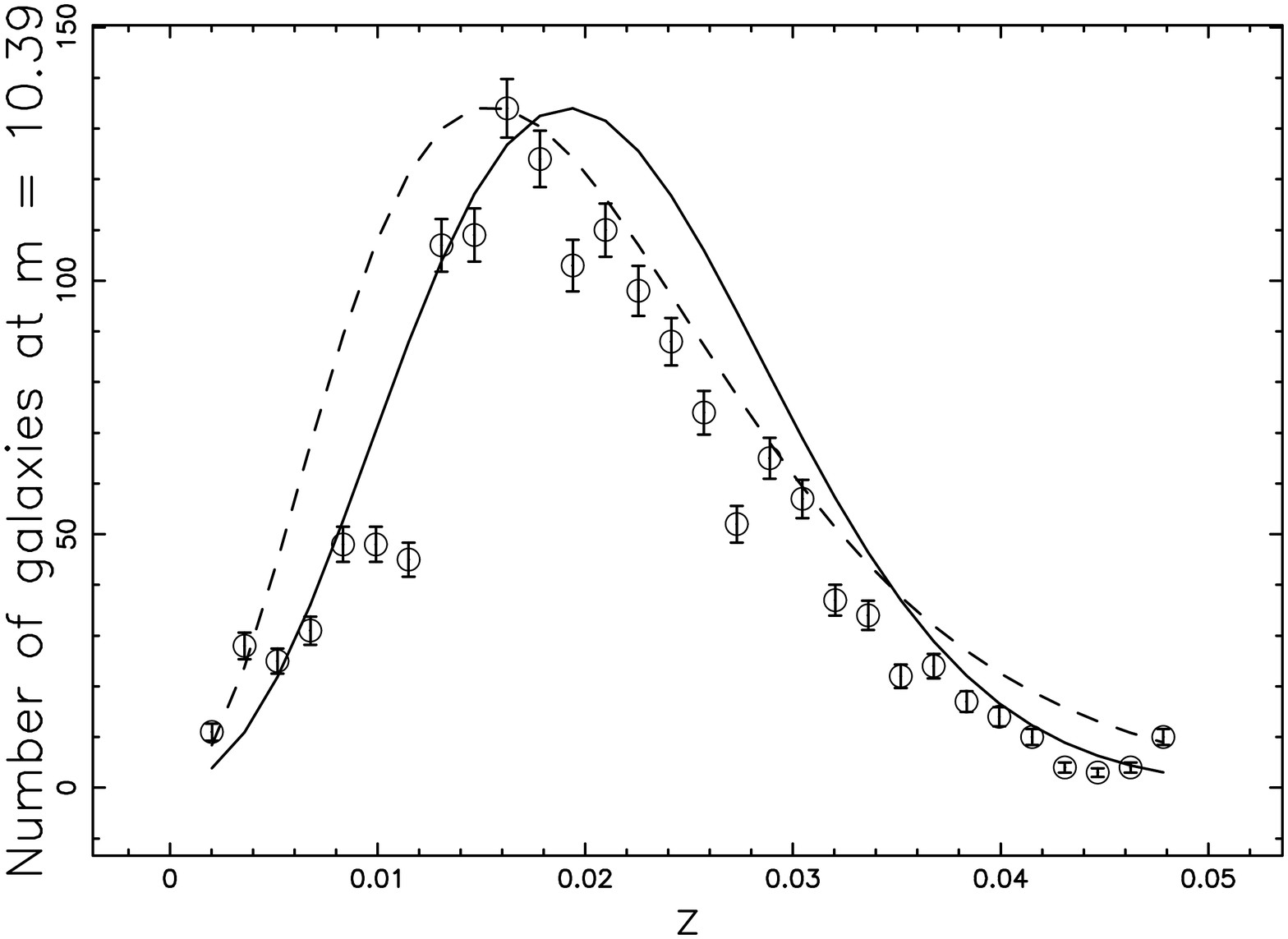}
\end {center}
\caption{
The galaxies  of the 2MRS with
$ 10.31     \leq  m   \leq 10.47  $  or
$ 1164793  \frac {L_{\sun}}{Mpc^2} \leq
f \leq  1346734 \frac {L_{\sun}}{Mpc^2}$
are  organized in frequencies versus
heliocentric  redshift,
(empty circles);
the error bar is given by the square root of the frequency.
The maximum frequency of observed galaxies is
at  $z=0.017 $.
The full line is the theoretical curve
generated by
$\frac{dN}{d\Omega dz df}(z)$
as given by the application of the Schechter LF
which  is equation~(43) 
in \cite{Zaninetti2019a}
and the dashed line
represents the Weibull  LF
which  is equation~(\ref{nfunctionzweibull}).
The Weibull LF parameters 
are   $c=1/2$ and
$M^*=-20.65$,   
$\chi^2= 198$  for the Schechter LF    and
$\chi^2= 452$  for the Weibull    LF.
}
          \label{weibull_photo_max}%
    \end{figure}

\begin{figure}
\begin{center}
\includegraphics[width=6cm]{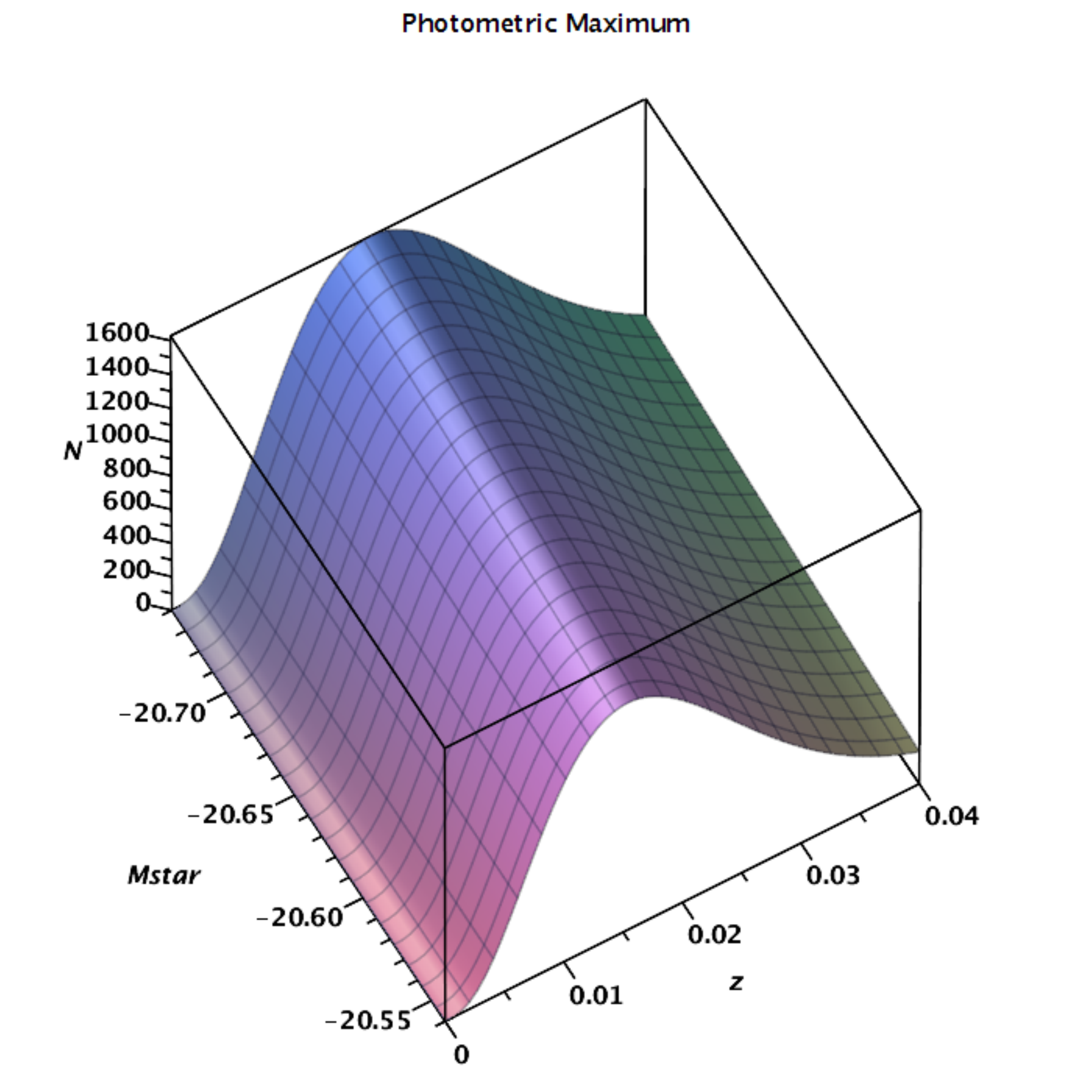}
\end {center}
\caption
{
The theoretical number of galaxies for 
the Weibull LF as function of the redshift 
and $M^*$;
parameters as in Figure~\ref{weibull_photo_max}.
}
          \label{maxz2d}%
    \end{figure}

A {\it third} approximate result is obtained 
using  a Taylor expansion of 
equation (\ref{eqnozmax}) around $z=2 z_{crit}$ of order 3,
which gives
\begin{eqnarray}
z_{max}=z_{crit} \times  
\nonumber \\
\frac
{
24\,{64}^{c}{c}^{3}-28\,{c}^{3}{16}^{c}+{2}^{2\,c+2}{c}^{3}-4\,{c}^{3}
{256}^{c}+4\,{64}^{c}{c}^{2}-12\,{c}^{2}{16}^{c}+{2}^{2\,c+2}{c}^{2}+c
{16}^{c}-c{4}^{c}-A-{4}^{c}
}
{
c \left( 12\,{64}^{c}{c}^{2}-2\,{c}^{2}{256}^{c}-14\,{c}^{2}{16}^{c}+2
\,{c}^{2}{4}^{c}+3\,c{64}^{c}-9\,c{16}^{c}+3\,c{4}^{c}-{16}^{c}+{4}^{c
} \right) 
}
\quad  ,
\end{eqnarray}
where 
\begin{eqnarray}
A=
\Big (
-4\,{c}^{4}{16}^{c}+40\,{64}^{c}{c}^{4}+32\,{1024}^{c}{c}^{4}+
56\,{64}^{c}{c}^{3}-48\,{c}^{3}{256}^{c}-8\,{c}^{3}{16}^{c}+14\,{64}^{
c}{c}^{2}-3\,{c}^{2}{16}^{c}
\nonumber  \\
+2\,c{16}^{c}
-3\,{c}^{2}{256}^{c}-2\,c{64}
^{c}+8\,{1024}^{c}{c}^{3}-60\,{256}^{c}{c}^{4}-4\,{4096}^{c}{c}^{4}+{
16}^{c}
\Big )^{1/2}
\quad  .
\end{eqnarray}
A graphical display of the Taylor 
solution is reported 
in Figure \ref{zmaxtaylor} as the red full line.
Figure \ref{weibull_photo_max}
reports the number of  observed  galaxies
for the  2MASS Redshift Survey (2MRS)  catalog  at   a given
apparent magnitude  and
both the Schechter and the Weibull models for the number 
of galaxies as functions of the redshift.
The influence on the above curve  of varying $M^*$ 
is reported in Figure~\ref{maxz2d}.

The mean redshift   for galaxies  
$ { \langle z \rangle } $
is 
\begin{equation}
 { \langle z \rangle }
=
\frac
{
\int_0^\infty  z \, \frac{dN}{d\Omega dz df} dz
}
{
\int_0^\infty  \frac{dN}{d\Omega dz df}  dz
}
\quad .
\end{equation}
The mean redshift
for the Weibull  LF
as  a function of $z_{crit}$
when $c=1/2$
is
\begin{equation}
 { \langle z \rangle } (z_{crit}) =
4\,z_{{{\it crit}}}
\quad when\quad c=1/2 \quad ,
\end{equation}
or  as a function of the flux
\begin{equation}
 { \langle z \rangle } (f) =
\frac
{
2\,\sqrt {\pi\,f{10}^{ 0.4\,{\it M_{\sun}}- 0.4\,{\it M^*}}}H_{{0}}
}
{
\pi\,fc_{{l}}
}
\quad when \quad c=1/2 \quad ,
\end{equation}
where $M_{\sun}=3.39$ is the reference magnitude
of the sun at the considered bandpass,
or  as a function of the apparent magnitude
\begin{equation}
 { \langle z \rangle } (m) =
\frac
{
4\,10^{-5}\sqrt {{{\rm e}^{ 0.921\,{\it M_{\sun}}-
 0.921\,m}}{10}^{ 0.4\,{\it M_{\sun}}- 0.4\,{\it M^*}}}H_{{0}}
}
{
{{\rm e}^{ 0.921\,{\it M_{\sun}}- 0.921\,m}}c_{{l}}
}
\quad when\quad c=1/2 \quad .
\end{equation}

\subsection{Mean absolute magnitude}

The absolute magnitude  which can be observed as a function of the
limiting apparent magnitude, $m_L$, is
\begin{equation}
M_L =
m_{{L}}-5\,{\it \log_{10}} \left( {\frac {{\it c}\,z}{H_{{0}}}}
 \right) -25
\quad ,
\label{absolutel}
\end{equation}
where $m_L$=11.75 for  the 2MRS catalog. 

\begin{figure*}
\begin{center}
\includegraphics[width=6cm]{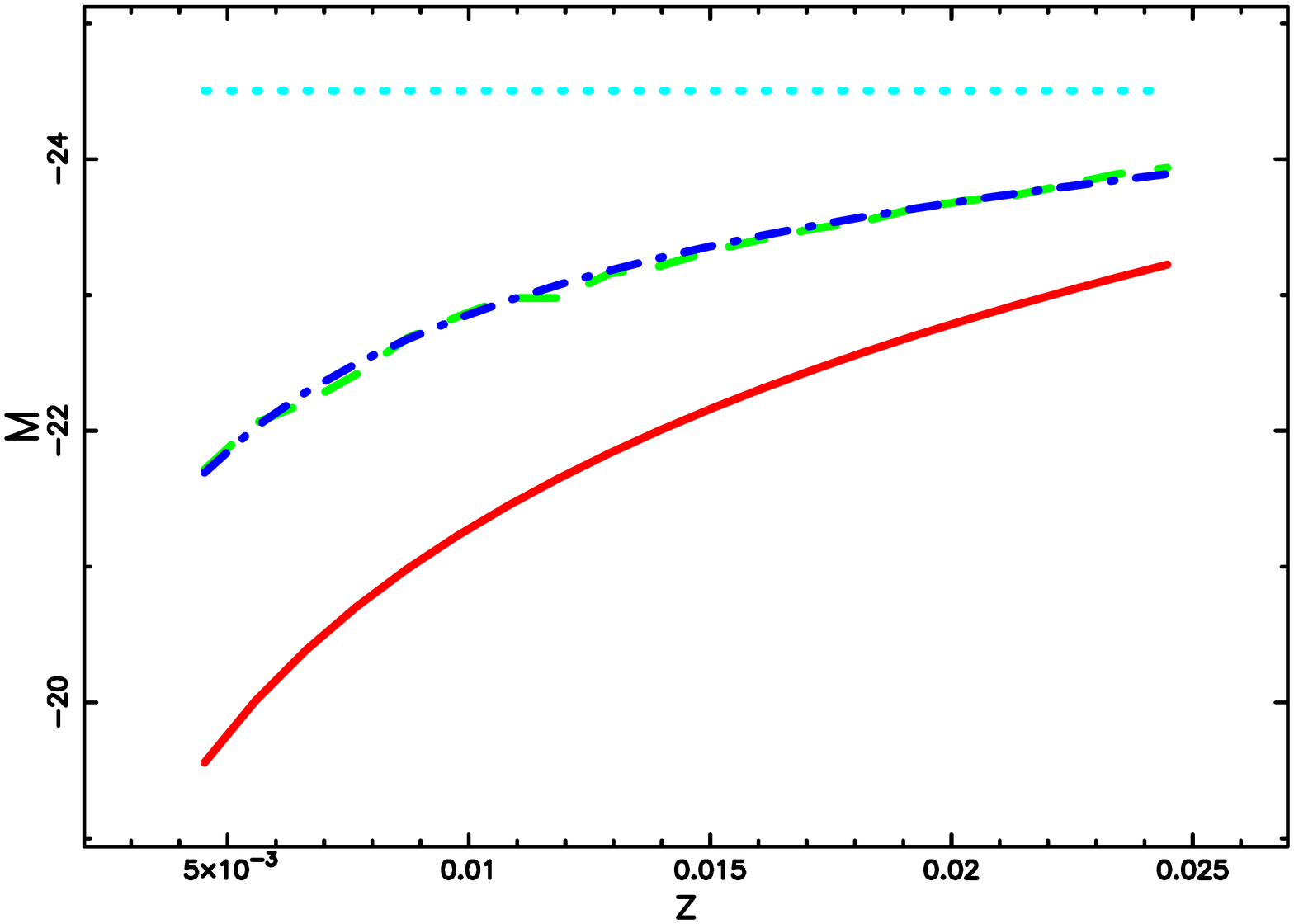}
\end {center}
\caption
{
Average absolute magnitude of the
galaxies belonging to the 2MRS
(green-dashed line),
theoretical average absolute magnitude
for the   truncated Weibull  LF
(blue dash-dot-dash-dot line)
as given by equation (\ref{meanabsolutetruncated}),
lower theoretical curve as represented by
Equation~(\ref{absolutel})
(red line)
and
minimum absolute magnitude observed (cyan dotted line).
}
 \label{weibull_bias}%
 \end{figure*}

The  theoretical average  absolute magnitude
of  the   truncated Weibull LF,
see  equation (\ref{meanabsolutetruncated}),
can be compared 
with the observed average absolute magnitude of the 2MRS
as a function of the redshift.
To fit the  data,
we assumed the following empirical  dependence on the redshift
for the characteristic   magnitude of the
truncated Weibull LF
\begin{equation}
M^* = -25.14 + 4
\Bigg (1- 
\Big ({ \frac{z-z_{min}} {z_{max}-z_{min}}}\Big)^{0.7} 
\Bigg)
\quad  .
\label{mstarz}
\end{equation}
This relationship models the decrease of the characteristic
absolute magnitude as a function of the redshift
and allows us to match the observational and theoretical data.
The lower bound  in absolute magnitude  is given
by the minimum magnitude of the selected bin,
the upper bound  is given by
equation (\ref {absolutel}),
the characteristic magnitude varies according
to  equation (\ref {mstarz})
and  Figure \ref{weibull_bias} reports
a comparison between the theoretical and the observed absolute magnitude
for the 2MRS catalog.

\section{Conclusions}

{\bf Truncated Weibull distribution}
We derived the PDF,
the DF,
the average value,
the $r$th moment,
the variance,
the median,
the mode,
an expression to generate  random numbers
and
the way to obtain  the two parameters, $b$ and $c$,
by the MLE
for the  truncated Weibull distribution.

{\bf Weibull luminosity function}
We derived the Weibull LF in the standard and the truncated case:
the application  to both
the SDSS Galaxies
and to the QSOs 
in the range of redshift $[0.3,0.5]$
yields 
a lower   reduced  merit function
compared  to  Schechter  LF, see 
Tables \ref{chi2valuelf} 
and
\ref{qsoweibullfit}.

{\bf Cosmological applications}

The number of galaxies as functions of the 
redshift, the flux and the solid angle 
for  the
Weibull LF in the pseudo-Euclidean universe
presents a maximum 
which can be compared with 
the observed one 
for   the 2MRS, see Figure 
\ref{weibull_photo_max}.
The truncated Weibull  LF produces a good fit
to  the average absolute magnitude of the
2MRS galaxies as a function of the redshift, 
see Figure \ref{weibull_bias}.

\providecommand{\newblock}{}

\end{document}